**Molecular beam epitaxy and defect structure of Ge (111)/epi-Gd$_2$O$_3$ (111) /Si (111) heterostructures**

Krista R Khiangte[1], Jaswant S Rathore[1], Sudipta Das[2], Ravinder S Pokharia[2], Jan Schmidt[3], H. J. Osten[3], Apurba Laha[2], Suddhasatta Mahapatra[1]

[1]*Department of Physics, Indian Institute of Technology Bombay, Mumbai, INDIA*

[2]*Department of Electrical Engineering, Indian Institute of Technology Bombay, Mumbai, INDIA*

[3]*Institute of Electronic Materials and Devices, Leibniz Universität Hannover, Schneiderberg 32, 30167 Hanover, Germany*

**Abstract**

Molecular beam epitaxy of Ge (111) thin films on epitaxial-Gd$_2$O$_3$/Si(111) substrates is reported, along with a systematic investigation of the evolution of Ge growth, and structural defects in the grown epilayer. While Ge growth begins in the Volmer-Weber growth mode, the resultant islands coalesce within the first ~ 10 nm of growth, beyond which a smooth two-dimensional surface evolves. Coalescence of the initially formed islands results in formation of rotation and reflection microtwins, which constitute a volume fraction of less than 1 %. It is also observed that while the stacking sequence of the (111) planes in the Ge epilayer is similar to that of the Si substrate, the (111) planes of the Gd$_2$O$_3$ epilayer are rotated by 180° about the [111] direction. In metal-semiconductor-metal schottky photodiodes fabricated with these all-epitaxial Ge-on-insulator (GeOI) samples, significant suppression of dark current is observed due to the presence of the Gd$_2$O$_3$ epilayer. These results are promising for application of these GeOI structures as virtual substrates, or for realization of high-speed group-IV photonic components.

**I. INTRODUCTION**

The germanium (Ge)-on-insulator (GeOI) technology is rapidly emerging as an enabling platform for a variety of applications in microelectronics, photonics, and photovoltaics. As the silicon (Si) microelectronics industry encounters technological and economic challenges to enhance speed and reduce power consumption, while minimizing the footprint of the field-effect transistors on microprocessor chips, mobility enhancement approaches are being aggressively pursued to meet performance requirements of the sub-22-nm technology node. In this context, GeOI is attracting tremendous attention as an engineered material system with high carrier mobilities, while harnessing the "on-insulator" advantages [1-4]. Besides, GeOI substrates are mechanically more stable and less expensive than those of bulk Ge, thus providing a practical advantage in device processing.

GeOI is also being investigated as a candidate system for high-speed photodetectors [5, 6] and templates for epitaxial growth of GaAs [7]. Since the thermal expansion coefficient and lattice constant of GaAs are almost identical to those of Ge, GeOI substrates offer the possibility of monolithic integration of GaAs onto the Si CMOS platform [8]. This would pave the way not only for the realization of electro-photonic integrated circuits (EPICs), but also for the cost-effective development of III-V semiconductor-heterosystems-based multi-junction solar cells, for the terrestrial photovoltaics market [9].

Several techniques have been investigated and established for preparation of GeOI substrates, the most notable amongst them being the Smart Cut$^{TM}$ technology, which involves transfer of Ge layers onto a Si/SiO$_2$ surface, from a Czochralski-grown Ge wafer [4, 10, 11]. Other techniques include Ge condensation [4, 12] and liquid phase epitaxy (LPE) [4]. While the Smart Cut$^{TM}$ technology is cost-intensive, the two other techniques suffer from high thermal budget and limited lateral extent of GeOI (~ 20 $\mu$m), respectively [4]. On the other hand, epitaxially-grown, single crystalline, high-$k$ dielectric oxides on Si have emerged in the recent past as attractive alternative templates for GeOI fabrication, with significant developments reported for a variety of perovskites [13] and lanthanide oxides (Ln$_2$O$_3$) [14]. In particular, epitaxial growth of yttria (Y$_2$O$_3$) [15], praseodymia (Pr$_2$O$_3$) [16-18], scandia (Sc$_2$O$_3$) [19], neodymia (Nd$_2$O$_3$) [20], and gadolinia (Gd$_2$O$_3$) [21-25] on Si(111) and Si(001) substrates have been investigated by different research groups. However, epitaxy of Ge on these epi-Ln$_2$O$_3$ insulating layers is yet to be thoroughly studied and optimized. To the best of our knowledge, Ge growth on only epi-Pr$_2$O$_3$/Si(111) has been comprehensively studied [26-30].

Here, we report on the molecular beam epitaxy (MBE) growth of Ge layers on Gd$_2$O$_3$/Si(111) and present a detailed study of their crystal quality and defect structure. Our results demonstrate that high quality (111)-oriented Ge epilayers, with a smooth surface, can be grown on Gd$_2$O$_3$/Si (111). We also report the reduction of dark current in photodiodes fabricated with these all-epitaxial GeOI samples.

## II. EXPERIMENTAL

The samples were grown in a RIBER MBE (Compact-12) chamber equipped with an in-situ reflection high energy electron diffraction (RHEED) system. The base pressure in the growth chamber was maintained at ~ $6.5 \times 10^{-10}$ mbar. The 100 mm epi-Gd$_2$O$_3$/Si(111) substrates used in this work were obtained from Leibniz Universität Hannover, where the thin (~ 10 nm) epitaxial Gd$_2$O$_3$ layers were also grown by MBE using a DCA S1000 multi-chamber system [23]. Details of the Gd$_2$O$_3$ epitaxy on Si(111) can be found elsewhere [22]. Here, we briefly explain only the relevant details of the oxide growth recipe. At first, a 10-nm-thick Si buffer layer is grown on HF-cleaned 100-mm Si(111) wafers at 750 °C. This leads to the (7x7) surface reconstruction, which can be clearly observed in RHEED. The substrate is then cooled down to 650 °C and the 10-nm-thick Gd$_2$O$_3$ layer is grown using an electron beam evaporator.

Additional molecular oxygen is supplied via a piezo-electric leak valve, at a partial pressure of $5 \times 10^{-7}$ mbar. This is known to significantly improve the dielectric properties of the epi-Gd$_2$O$_3$ layer [31].

For the Ge growth, the Gd$_2$O$_3$/Si(111) substrates (shipped from Hannover to Mumbai) were first baked for 10 hours at 150 °C, within the load-lock chamber of the MBE system. Subsequently, the substrates were transferred to the ultrahigh vacuum (UHV) growth chamber and further heated to 700 °C, to obtain a clean and smooth epi-Gd$_2$O$_3$ surface. Ge epilayers, approximately 360 nm thick, were grown at $T_G = 550$ °C, at a growth rate of 1.6 nm min$^{-1}$.

In-situ growth monitoring was done by RHEED (Staib Instruments Inc.), at a beam energy of 12 keV. High-resolution X-ray diffraction (HRXRD) scans and pole figures were recorded in a Rigaku Smartlab instrument, which is capable of performing scans in both out-of-plane and in-plane geometries. The diffractometer is equipped with a 9 kW rotating Cu anode, a parabolic mirror, a double-crystal Ge (220) monochromator, and several Soller slits. While, wide ω-2θ scans were performed in the high resolution mode, wherein the two-bounce monochromator was used on the incident side, the θ-2θ and the pole figure measurements were carried out in a low resolution mode, with a 5.0 degree Soller slit replacing the monochromator.

For the pole figure measurements, the polar angle (α) was varied from 0 ° (in which the scattering vector is in the sample plane) to 90 ° (in which the scattering vector is parallel to the surface normal) by varying the sample tilt (χ-rotation), while the azimuthal angle (β) was varied by sample rotation (Φ - rotation) about the surface normal. Additionally, θ-2θ scans have also been recorded for specific (α, β) spots of the pole figure. The microstructure, stacking configurations, and defect structure were also analysed by recording cross-sectional high-resolution transmission electron microscopy (HRTEM) images along the $[1\bar{1}0]$ direction, using a JEOL 200 microscope, operating at voltages up to 200 kV. For the fabrication of metal-semiconductor-metal (MSM) back-to-back Schottky diodes, a 350-nm-thick SiO$_2$ passivation layer was grown at 200 °C, atop the epi-Ge layer, by inductively-coupled-plasma-assisted chemical-vapour-deposition (ICPCVD). Subsequently, inter-digitated electrodes (IDE) were patterned by single level optical lithography, SiO$_2$ etching, and deposition of Ni (20 nm)/Au (50 nm) contacts by electron beam evaporation. The current-voltage characteristics of the MSM structures were measured at room temperature with an Agilent B1500A Semiconductor Device Parameter Analyser.

**III. RESULTS AND DISCUSSION**

RHEED images recorded at different stages of Ge growth are shown in Figure 1. The image of Figure 1(a), recorded immediately before the commencement of Ge growth, reveals a streaky pattern with good signal-to-noise ratio and minimum intensity modulation along the diffraction rods. This indicates that the Gd$_2$O$_3$/Si(111) starting surface is clean and flat. When Ge growth begins, the streaky pattern persists for approximately 35 s (corresponding to ~ 1 nm of Ge), although the signal-to-noise ratio worsens (Figure 1(b)). This may be attributed to the formation of an amorphous Ge-oxide layer, similar to that reported for Ge growth on PrO$_2$/Si(111) [26].

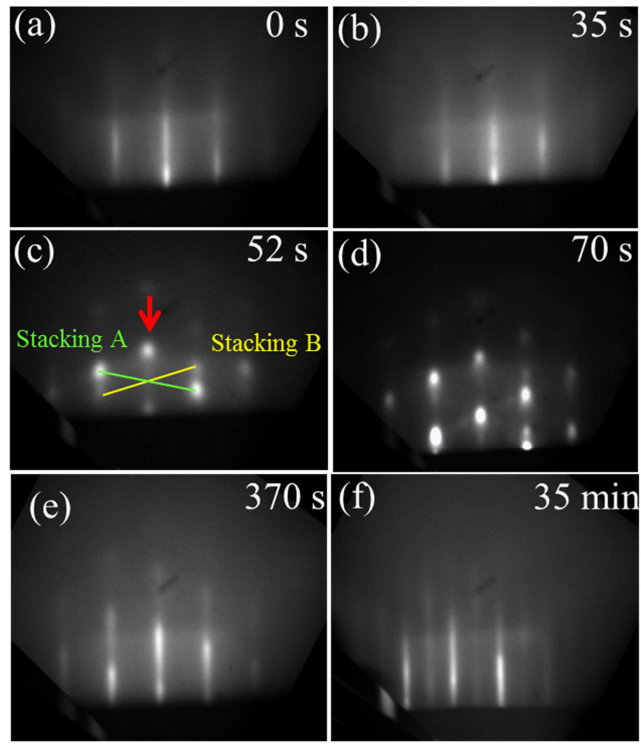

*Figure 1: RHEED images recorded at different stages of Ge growth: (a) Image of the Gd$_2$O$_3$(111) surface just before the start of Ge growth. Images recorded (b) 35 s, (c) 52 s, (d) 70 s, (e) 370 s, and (e) 35 min after commencement of Ge growth.*

After ~ 52 s of Ge growth, a substantial change in the RHEED pattern occurs, whereby a faint spotty pattern appears, while the low signal-to-noise ratio continues to persist (Figure 1(c)). This spotty pattern suggests the onset of 3D island formation, as typical for Volmer-Weber (V-W) growth, and as has been previously observed in Ref. [26, 32]. A closer look at Fig. 1(c) reveals a superposition of two spotty patterns of significantly different intensities. The patterns may be made to coincide with each other if one of them is mirrored about the central rod (marked by an arrow). The observed superposition of the two spotty patterns may be attributed to the co-existence of type-A and type-B stacked Ge islands at the initial stages of growth. Here, type-A refers to the stacking of (111)-oriented planes similar to that of the Si substrate, while type-B refers to stacking of the same planes rotated by 180° around the surface normal.

After ~ 70 s of growth (corresponding to the deposition of ~ 2 nm of Ge), the intensity of one of the (two superimposing) spotty patterns is strongly enhanced (Figure 1(d)), suggesting preferential growth of islands with one stacking sequence (type-A), over those with the other stacking sequence (type-B).

Figure 1(e) reveals that the "spotty" nature of the RHEED intensity distribution almost completely disappears after 370 s of growth (~ 10 nm Ge deposition). On further continuation of Ge growth, a fully streaky RHEED pattern develops, as shown in Figure 1(f). The low intensity streaks in this RHEED image, between the bulk Ge streaks, reflects the c(2×8) surface reconstruction of the epi-Ge(111) surface [33, 34]. From the above discussion of the RHEED images, it may therefore be concluded that while Ge epitaxy on $Gd_2O_3$ (111)/Si(111) initiates in the V-W growth mode, two-dimensional layer-by-layer growth is recovered after Ge deposition equivalent to a layer thickness of ~ 10 nm.

The ω-2θ XRD-scan of Figure 2(a), taken over the angular range of 22° to 100°, establishes the formation of single-crystalline, fully (111)-oriented, Ge epilayers, wherein only the (111) and the (333) reflections of Ge are visible, at 27.30° and 90.11°, respectively. The Bragg positions of the Ge peaks yield the bulk value for the interplanar spacing of the {111} planes, which implies that the grown epi-Ge layer is fully relaxed. A comparison with Figure 2(b), which displays the wide ω-2θ X-ray diffractogram of the $Gd_2O_3$(111)/Si(111) substrate, reveals that no significant structural modification of the $Gd_2O_3$(111) layer takes place, either during the initial surface-preparation step or during the subsequent Ge growth. In both Figs., the Si (111) and (222) peaks are observed to coincide with the (222) and (444) peaks of the epi-$Gd_2O_3$ layer, respectively.

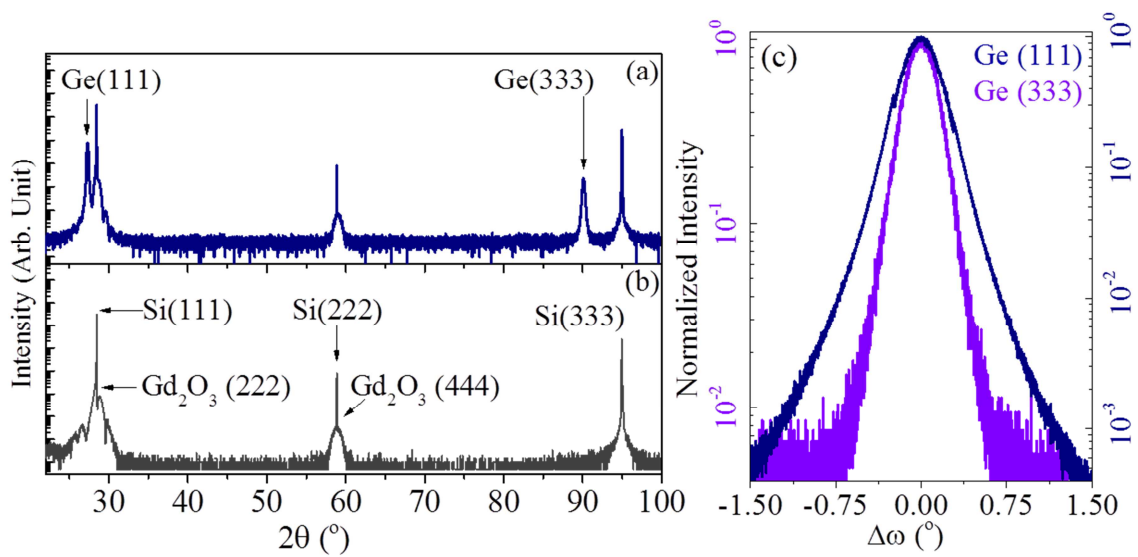

*Figure 2: (a) wide ω-2θ scan for (a) the Ge(111)/$Gd_2O_3$(111)/Si(111) epilayers and (b) for the $Gd_2O_3$ (111)/Si(111)substrate.(c) ω-rocking curve diffractograms for the two Ge reflections.*

Since the bulk lattice constant of $Gd_2O_3$ $(a_{Gd_2O_3} = 10.812 \text{ Å})$ is only slightly smaller than twice that of Si $(a_{Si} = 5.431 \text{ Å})$, the 2nd, 4th and 6th order reflection of the $Gd_2O_3$ (111) planes are superimposed on the 1st, 2nd, and 3rd order reflection of the of Si (111) planes, respectively. The calculated out of plane lattice spacing of $Gd_2O_3$ $(d_{111})$ in the thin oxide layer was found to be larger than the corresponding bulk value, revealing that the oxide layer is compressively stressed. This situation, where the Ge epilayer is fully relaxed, while the $Ln_2O_3$ layer is compressively stressed, was also observed in Ref. [27] and [35], for Ge(111)/epi-$Pr_2O_3$(111)/Si(111) heterostructures.

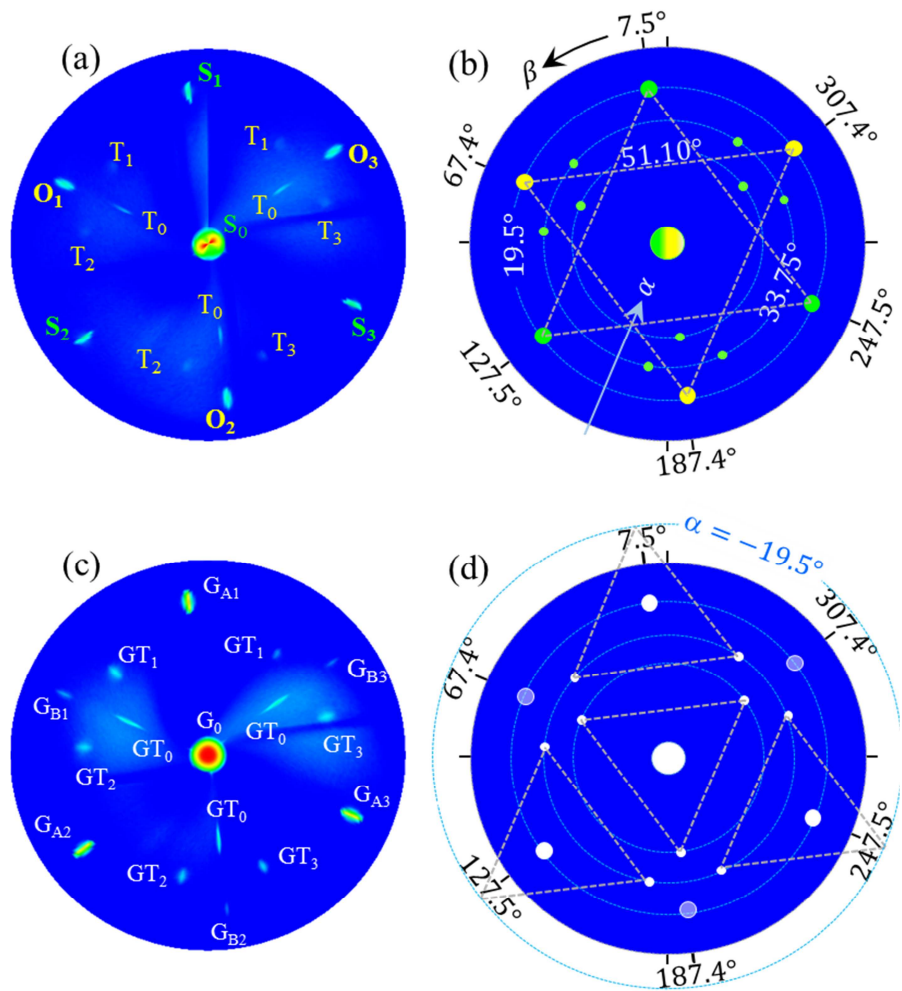

*Figure 3: XRD - pole figure measurement at the Bragg angle corresponding to the (a) Si (111) and (c) Ge (111) reflections. (b) and (d) are schematic representations of the pole figure images of (a) and (c) respectively, illustrating the important α contours and β positions, as well as elucidating the different triads of points.*

A qualitative estimate of the crystal quality of the epi-Ge layers was obtained by measuring the full-width-at-half-maximum (FWHM) of ω-diffractograms, corresponding to both Ge(111) and Ge(333) reflections, as shown in Fig. 2(c). The measured values, 0.38° and 0.33° respectively, together with the fact that the (333) peak is narrower than the (111) peak, suggests negligible tilting of the (111) planes of the epilayer, w.r.t. the surface [36].

To probe the nature and structure of defects in the Ge(111)/epi-$Gd_2O_3$(111)/Si(111) heterostructures on a global scale, systematic pole figure measurements were performed (Figure 3). Figures 3(a) and 3(c) show the Si(111) and Ge(111) pole figures recorded at 2θ = 28.44° and 2θ = 27.28°, respectively. Figures 3(b) and 3(d) are the corresponding schematic representations of the pole figures, depicting the polar ($\alpha$) and the azimuthal ($\beta$) angles for the different reflections. A pole figure recorded at 2θ = 28.44° for a bare epi-$Gd_2O_3$(111)/Si(111) substrate (not shown) revealed features exactly similar to that shown in Fig. 3(a), indicating that no significant crystallographic change occurs to the oxide layer during Ge epitaxy. This finding is in agreement with that of the ω-2θ diffractograms, discussed earlier in the context of Figure 2. The four spots, $S_0$ ($\alpha = 90°$, $\beta = 0°$), $S_1$ ($\alpha = 19.5°$, $\beta = 7.5°$), $S_2$ ($\alpha = 19.5°$, $\beta = 127.5°$), and $S_3$ ($\alpha = 19.5°$, $\beta = 247.5°$) (marked with green circles in Fig. 3(b)) correspond to the (111), ($\bar{1}11$), ($1\bar{1}1$), and ($11\bar{1}$) crystal planes of the Si substrate, reflecting the threefold symmetry of the [111] axis. However, at the same value of $\alpha$, another triad of spots, $O_1$ ($\alpha = 19.5°$, $\beta = 67.4°$), $O_2$ ($\alpha = 19.5°$, $\beta = 187.4°$), and $O_3$ ($\alpha = 19.5°$, $\beta = 307.4°$) is also visible in Figs. 3(a) (marked with yellow circles in Fig. 3(b)).

This triad, wherein the neighbouring spots are separated by the same azimuthal angle ($\Delta\beta = 120°$) as that of the ($S_1$, $S_2$, $S_3$) triad, corresponds to the ($\bar{2}22$), ($2\bar{2}2$), and ($22\bar{2}$) crystal planes of the $Gd_2O_3$ epilayer. The fact that the ($O_1$, $O_2$, $O_3$) triad is visible at the Si(111) Bragg angle is explained by the overlap of the 2$n$-order reflections of the $Gd_2O_3$ (111) planes with the $n$-th order reflections of Si (111) planes. Thus, $O_1$, $O_2$, and $O_3$, together with $S_0$, reflect the cubic symmetry of the epi-$Gd_2O_3$ layer. Here, it is important to note that the ($O_1$, $O_2$, $O_3$) triad is rotated by 180° w.r.t. the ($S_1$, $S_2$, $S_3$) triad (See Fig. 3(b)). Additionally, it was verified that only Si{$\bar{1}11$} reflections contribute to the ($S_1$, $S_2$, $S_3$) triad (See discussions in the context of Fig. 4(a)). These two observations indicate that the stacking sequence of the $Gd_2O_3$ (111) layer, w.r.t. the Si(111) substrate, is of type-B on a global scale. The type-B stacking may be also interpreted as reflection of the {$\bar{1}11$} planes about the (111) mirror plane. This is evident in Fig. 3(b), where it is easily seen that the $O_n$ ($n = 1, 2, 3$) spots can be generated by inverting the $S_n$ ($n = 1, 2, 3$) points about $S_0$. A direct evidence of mirroring of the ($11\bar{1}$) planes about the (111) interface plane is provided by HRTEM images, shown later in Figure 5(b).

Turning to Fig. 3(b) (along with the schematic of Fig. 3(d)), which is recorded at the Ge(111) Bragg angle, we find that a hexad of spots (marked by white circles in Fig. 3(d)) appear for $\alpha = 19.5°$, at exactly the same azimuthal positions as that of the ($S_1$, $S_2$, $S_3$) and the ($O_1$, $O_2$, $O_3$) triads of Figs. 3(a).

However, it is important to note that the Ge(111) spots coinciding with the ($S_1$, $S_2$, $S_3$) positions, i.e. ($G_{A1}$, $G_{A2}$, $G_{A3}$), are much more intense than those coinciding with the ($O_1$, $O_2$, $O_3$) positions, i.e. ($G_{B1}$, $G_{B2}$, $G_{B3}$). This reflects that the epi-Ge(111) layer is predominantly type-A stacked w.r.t. the Si(111) substrate, albeit a small fraction of type-B-stacked region also exists. Since the ($G_{B1}$, $G_{B2}$, $G_{B3}$) triad appears at the same value of $\alpha$ as that of the ($G_{A1}$, $G_{A2}$, $G_{A3}$) triad, it can be concluded that the type-B region represents rotational twins, formed due to 180 ° azimuthal rotation of the (111) planes, about the [111] twin axis. The situation is similar to that of the epi-Gd$_2$O$_3$ (w.r.t. the Si(111) substrate) but in case of the Ge(111) epilayer, the type-B regions constitute a very small volume fraction (See discussion below), restricted within the rotational microtwins.

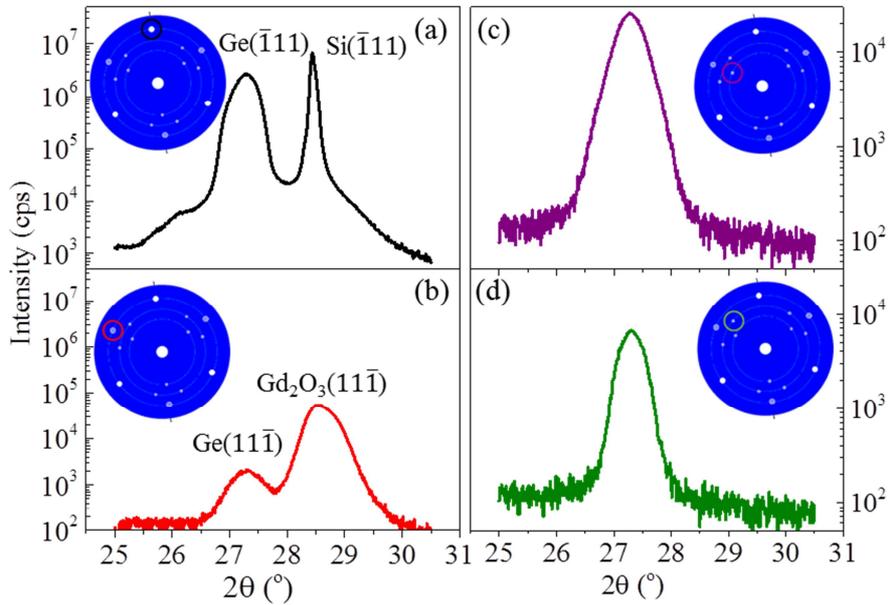

*Figure 4: θ-2θ diffractograms for (a) $G_{A1}$, (b) $G_{B1}$, (c) $G_{T0}$ and (d) $G_{T1}$ spots, which are shown in the corresponding insets by the encircled points.*

In addition to the rotational microtwins, reflection microtwins are also observed in the epi-Ge layer. The triad of spots in the pole figure of Fig. 3(c) indexed as GT$_0$ appear at $\alpha = 51.1°$. This means that these spots correspond to reflection from Ge{111} crystal planes which are tilted by 38.9 °, w. r. t. the global Ge (111) planes. As studied in detail by Niu et. al [37], the GT$_0$ triad appears due to reflection twins of the (111) plane, which are reflected about the three {$\bar{1}$11} planes. The other three triads, GT$_1$, GT$_2$, and GT$_3$, for which one of the apices lies outside the detection range of the pole figure ($\alpha = -19.5°$), are symmetrically equivalent to the GT$_0$ triad (See Fig. 3(d)). For example, the GT$_1$ triad appears due to reflection twins of the ($\bar{1}11$) plane, reflected about the three planes inclined at 70.5° and sharing edges with it, i.e. ($\bar{1}\bar{1}1$), ($\bar{1}1\bar{1}$), and (111).

Similar pole-figure spots are also visible in Fig. 3(a), indexed as $T_0$ ($\alpha = 51.1°$) and $T_n$ (n = 1, 2, 3 and $\alpha = 33.75°$). These spots may not be attributed to any twinning of the epi-$Gd_2O_3$ layer, since their azimuthal positions indicate to twinning of A-type-stacked {111} planes. We believe that these spots appear due to micro-twinned regions in the epi-Si layer, underneath the oxide. In Ref. [30], pole figure spots due to reflection microtwins were not observed at the Si (111) Bragg angle, although they were visible at the Ge (111) Bragg angle. It may therefore be inferred that the microtwins of the Ge epilayer observed in our samples are not correlated to those of the epi-Si layer.

To obtain a quantitative estimate of the ratio of the (global) A-type-stacked and B-typed-stacked regions (within the rotation twins) of the Ge(111) epilayer, θ-2θ diffractograms were recorded for the range $24° < 2\theta < 31°$, for pole figure spots at $G_{A1}$ ($\alpha = 19.5°$, $\beta = 7.5°$) and $G_{B1}$ ($\alpha = 19.5°$, $\beta = 67.4°$), or equivalently $S_1$ and $O_1$, respectively (Figure 4(a) and 4(b)). While the two peaks in Fig. 4(a), at $2\theta = 27.30°$ and $28.44°$, are due to the ($\bar{1}11$) reflections of Ge and Si, respectively, the two peaks in Fig. 4(b), at the same 2θ positions, are due to the Ge($11\bar{1}$) and $Gd_2O_3$($22\bar{2}$) reflections, respectively. The fraction of type-B Ge within the rotation twins, estimated by calculating the ratio of the Ge(111) peak areas at $G_{A1}$ and $G_{B1}$, yields only 0.8%, i.e. an amount equivalent to ~ 3 nm of Ge. The absence of the $Gd_2O_3$($22\bar{2}$) peak in Fig. 4(a) confirms that the epi-$Gd_2O_3$ layer is completely type-B stacked.

Figure 4(c) and 4(d) show the θ-2θ diffractograms recorded at $GT_0$ and $GT_1$ in the range $24° < 2\theta < 31°$. The Ge(111) peaks at the two spots (at $2\theta = 27.3°$) are of similar intensity, which in turn, are three-orders-of-magnitude weaker than the Ge(111) peak at $G_{A1}$ (or $G_0$). It may therefore be assumed that the fraction of the reflection twins is comparable to that of the rotation twins, estimated earlier. While the analysis of the previous section suggested the existence of micro-twinned regions in the epi-Si layer, no Si peak is seen in Figs. 4(c) and 4(d). This may be attributed to the fact that the diffracted intensity due to a small fraction of micro-twinned region in a 10-nm-thin epi-Si layer, buried beneath the 360-nm-thick Ge epilayer (and also the 10-nm-thick epi-$Gd_2O_3$ layer) is weaker than the background intensity of the measured diffractograms. However, the Si peak could be resolved in diffractograms similar to Figs. 4(c) and 4(d), recorded from the bare $Gd_2O_3$/Si(111) substrate (not shown).

Microstructure and defect analyses were further carried out by cross sectional HRTEM. The image of Figure 5(a), recorded over a large area of the sample, shows a high density of stacking faults (SF)/twins, and threading dislocations. The stacking fault density is lower in the upper half of the Ge–epilayer, suggesting that a virtual Ge(111) substrate of high crystalline quality may be obtained by growing thicker (~ 1 µm) epilayers, as also observed in the case of Ge(111)/epi-Pr$_2$O$_3$(111)/Si(111) epitaxy [30]. The SF/twins are observed to propagate at 70° w.r.t. the surface, which is consistent with the observations for Ge epitaxy directly on Si(111) substrates [38].

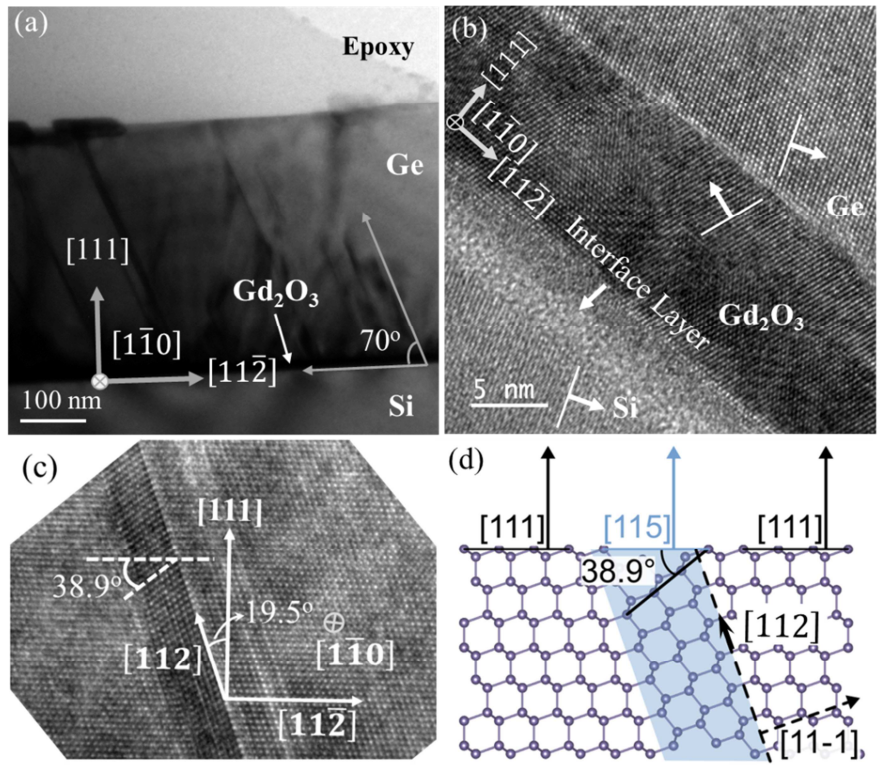

*Figure 5: (a) Cross sectional HRTEM image of the Ge(111)/Gd$_2$O$_3$(111) /Si(111) sample. (b) A high resolution image of the interface region, showing the stacking orientation of the Gd$_2$O$_3$ and the Ge epilayers, relative to that of the Si(111) substrate.(c) A close-up image of a reflection microtwin and (d) schematic visualization of twin lamella* [39].

Figure 5(b) shows the high resolution image of the Ge/epi-Gd$_2$O$_3$/Si interface. The type-A and type-B stacking of the epi-Ge and epi-Gd$_2$O$_3$ layers, respectively, w.r.t. to the Si substrate, is clearly visible in this image. The normal to the $(11\bar{1})$ planes are indicated for each region (Si, Gd$_2$O$_3$, and Ge) by arrows, which reveal that the stacking order of Si and Ge is the same. However, the stacking order of the Gd$_2$O$_3$ $(11\bar{1})$ planes is obtained by reflection of the Si $(11\bar{1})$ planes about the interface, as previously described in the pole figure analysis. Figure 5(b) also shows the formation of an interfacial layer at the Gd$_2$O$_3$/Si interface.

This layer, which appears to be amorphous in nature, is possibly formed during annealing at 700 °C, while preparing the epi-Gd$_2$O$_3$ surface for Ge epitaxy. A close-up image of a SF/twin region is shown in Figure 5(c). It is observed that the (111) crystal planes within the lamellar region are inclined w.r.t the (111) planes of the surrounding un-faulted Ge epilayer, by an angle of 38.9°. This lamellar region is thus [115]-oriented, as shown in the schematic of Figure 5 (d) [39]. The lamella results due to reflection of the un-faulted (111) planes, about the $(11\bar{1})$ plane (shown by the dotted line). In accordance with the pole figure analysis, it is thus evident that these regions represent the reflection microtwins. Although rotational twins could not be imaged in HRTEM, their presence is confirmed by the pole figure analysis, presented earlier.

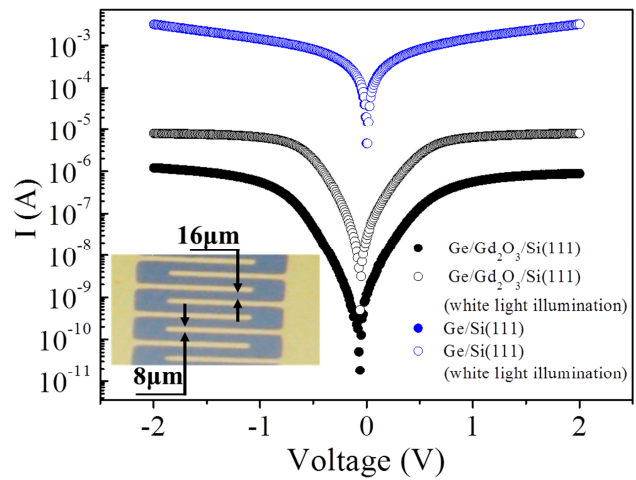

*Figure 6: I-V characteristics of MSM photodiodes fabricated with the Ge(111)/Gd$_2$O$_3$(111) /Si(111) (black) compared to that of MSM photodiodes fabricated with Ge/Si(001) epilayers(blue), with and without white light illumination. The inset shows a light microscope image of the interdigitated electrodes of the photodiode.*

One of the important "on-insulator advantages" is the suppression of leakage currents by isolating the (Si) substrate from the Ge epilayer. To verify this, we fabricated MSM back-to-back Schottky diodes on similar Ge/Gd$_2$O$_3$/Si(111) heterostructures, surface-passivated with a 350 nm thick SiO$_2$ capping layer. An optical image of the interdigitated contacts of the MSM structure is shown in the inset of Figure 6.

In the main panel, the current-voltage (I-V) characteristics of this diode is compared with the I-V characteristics of a similar MSM structure fabricated with a 500-nm-thick Ge(001) epilayer (with a 350 nm SiO$_2$ passivation layer atop) , grown directly on a Si(001) substrate. The details of the Ge(001)/Si(001) growth can be found elsewhere [40].

For both samples, the Ge epilayer was grown at $T_G$ = 450 °C. It is evident that the dark current is strongly supressed (by four orders of magnitude at bias voltage $\geq$ 1 V) for the diodes fabricated on the all-epitaxial Ge(111)/Gd$_2$O(111)/Si(111) heterostructures, in comparison to those fabricated on Ge(001)/Si(001) epilayers. On illumination with white light (with a DC Xenon arc lamp), no change is observed in case of the latter, while the photocurrent is measured to be 10 times higher (at bias voltage $\geq 1$ V) than the dark current, in case of the former. This observation is encouraging for development of high speed photodetectors on such all-epitaxial GeOI wafers, based on Ge or other group-IV alloys, such as $Ge_{1-x}Sn_x$ [41-43]

## IV. CONCLUSIONS

In conclusion, we reported the epitaxial growth of Ge (111) thin films by MBE on epi-$Gd_2O_3$ (111)/Si(111) substrates, and presented a detailed study of the crystal structure and nature of structural defects in such GeOI samples. The results also shed light on the nature of the Ge epitaxy, revealing island formation at the initial stages of growth, while coalescence of these islands and formation of a smooth surface beyond a layer thickness of ~ 10 nm. The suppression of stacking faults and twinning beyond a thickness of ~ 300 nm suggests that thick (~ 1 $\mu$m) epi-Ge (111)/ $Gd_2O_3$ (111)/Si (111) layers may be promising as virtual substrates for subsequent epitaxy, and also for enhanced performance of group-IV based photonic devices.

**ACKNOWLEGMENTS**

This research was funded by the Science and Engineering Research Board, Department of Science and Technology (DST), Government of India. KRK, JR, SD, RP, AL, and SM acknowledge support from the Centre of Excellence in Nanoeletronics (CEN) and Industrial Research and Consulting Centre (IRCC), Indian Institute of Technology Bombay.